# Quantum Spin Relaxation with THz Attempt Frequency in the 1/3-Fire, 2/3-Ice Ferrimagnet $SmMn_2Ge_2$

M. L. McLanahan and A. P. Ramirez

*Physics Department, University of California Santa Cruz, Santa Cruz, CA, 95064*

Ferrimagnets in which half the spins are disordered and half ordered, so-called half-fire, half-ice state, have been advanced theoretically as a route towards realizing exotic magnetic ground states. Here we demonstrate novel single-spin resonance in $SmMn_2Ge_2$, a 1/3-fire, 2/3-ice material. Using ac-susceptibility in the frequency range 0.1 Hz – 1 kHz we find a Debye-like resonance with Orbach relaxation temperature dependence, yielding an energy barrier $E_B \approx 840\ k_B$ and attempt frequencies in the THz range. These quantities strongly suggest single-spin relaxation among paramagnetic Sm spins where the flipping field is provided by the FM-ordered Mn spins at domain boundaries, a direct outcome of the 1/3-fire, 2/3-ice state.

## Introduction

The interest in spin-based information control remains high due to the prospect of fast, energy efficient devices. Many different materials platforms have been proposed for such devices [1-4] with a large range of potentially useful magnetic phenomena aided by a variety of structure types and magnetic elements, the combinatoric interplay of which can create conditions with competing interactions and resultant low energy excitations, important for devices. A magnetic platform of recent interest are Single-Molecule Magnets (SMMs), with device ambitions towards molecular-sized storage and quantum computing [5]. Lanthanide SMMs [6, 7], especially $Dy^{3+}$ based systems [8-11], in which magnetic relaxation seen with ac-susceptibility ($\chi_{ac}$) shows that the single ion anisotropy of rare-earths leads to large thermal energy barriers, a requirement for non-volatile memory.

Another potential platform for magnet-based logic is based on skyrmions, which are localized regions of spin reversal involving many spins. Ultimately one may want to control a single spin, but this goal presents a fundamental problem. Maintaining polarization of a single spin near room temperature requires a large field, $H \cong k_B T / g\mu_B$, which exceeds 200 Tesla. Of course ferromagnets achieve spin polarization at room temperature via the combined effects of the exchange field and multiple nearest neighbors, but such spins will not easily be flipped due to their mutual interactions. Thus, as a first step, it is desirable for the logic spins to sit on a different sublattice from the spins supplying the large polarizing field. This scenario is realized in so-called half-ice, half-fire ferrimagnets in which one of the sublattices is ordered (ice) and the other disordered (fire) [12]. Attractive materials platforms for such systems are realized in compounds $A_x B_y C_z$ where $C$ is an anion and the $A$ and $B$ cations are $4f$ and $3d$ elements. Because $4f$ and $3d$-derived spins possess vastly different interaction energies, the temperature can be tuned to lie between the two scales. An example of such a compound family is the rare-earth orthoferrites, $R$FeO$_3$ which exhibit large temperature regions below the Fe-driven Curie temperature (> 500 K) [13] where the $4f$ ions are essentially paramagnetic, albeit subject to an internal field.

A particularly interesting fire/ice materials platform with formula $A_x B_y C_z$ is the quasi-2D ThCr$_2$Si$_2$ structure-type, where the $A$-ions and $B$-ions form alternating layers, albeit with different spin densities. Lamellar structures are also desirable from a device perspective, and the $R$Mn$_2$Ge$_2$ series develops either an antiferromagnetic or a (canted) FM state below 300 K, depending on the size of the rare-earth ($R$) element [14]. We trained our focus on $SmMn_2Ge_2$, the compound sitting at the boundary between AF and FM states because $SmMn_2Ge_2$ develops a re-entrant FM phase among Mn ions below 100 K (the "ice" spins) and because the thermal energy is well matched to our frequency range [14-20].

To understand the interplay of Sm and Mn moments in $SmMn_2Ge_2$, we measured its ac-susceptibility, $\chi_{ac}$, motivated by the temperature-hysteretic behavior observed by Hou et al. [18] in the re-entrant region. Such hysteresis is evocative of spin glass behavior in a heavily disordered system but in a FM system with few defects it can be associated with domain wall pinning. Our $\chi_{ac}$ measurements show large, Debye-like, resonances in the re-entrant FM phase at 30 – 50 K and 0.1 – 1000 Hz. While Debye-like resonances have been observed in other FMs, in crystalline systems they usually reveal attempt frequencies in the $10^6$ Hz range, reflecting the physics of pinned domain walls [21]. By contrast, the Arrhenius behavior of the resonances in $SmMn_2Ge_2$ yields an attempt frequency, $\tau_0^{-1}$, in THz range, along with an energy barrier $E_B \approx 840\ k_B$. Taken together, these two quantities strongly suggest that the resonances involve single $Sm^{3+}$ spins, and their decreasing magnitude with field further suggests that the flipping field is supplied by the FM Mn sublattice via domain wall motion.

The samples used in this study were obtained from the UCSC sample archive [22]. The crystals were typically a few mm in their long direction which is in the *a-b* plane of the crystal. X-ray diffraction was deemed unnecessary because the uniqueness of $SmMn_2Ge_2$'s physical properties serves both to identify the compound as well as confirm the samples' high quality, e.g. by the sharpness of the phase transitions (see Supplemental Information). Both dc-magnetization and ac-susceptibility data were obtained using a Quantum Design MPMS3 magnetometer, with typical ac-drive fields of 1.0 Oe. Most of the data were obtained at zero static applied field, after quenching the main magnet. Samples from different batches were measured to validate reproducibility but most of the data presented were obtained on a sample of size 1.8 mm × 1.1 mm × 0.2 mm with the short axis aligned with the *c*-direction. This sample was measured with ac-drive field in the *a-b* plane, where demagnetization effects are minimized. Demagnetization corrections to the ac-susceptibility were not performed, however, due to the suspected inhomogeneity of the internal field responsible for the ac-signal, which is discussed in our analysis. The dc-susceptibility (see supplemental information [23]) agrees well with previously reported results for $SmMn_2Ge_2$ [14, 16, 18]. In $SmMn_2Ge_2$, the Mn ions form *a-b* plane layers with the smallest Mn-Mn distance being 2.92 Å separated by layers of less dense Sm ions with Sm-Sm distance of 4.134 Å. Because the Mn and Sm spins are ordered and not ordered, respectively in our measurement range, we refer to them as 2/3-ice and 1/3-fire. The Mn ions undergo a ferromagnetic (FM) transition at 350 K to a non-collinear state [20] in which each Mn plane has a net moment in the *c*-direction and is aligned with that of the other planes. On cooling to 154 K, the Mn planes become anti-aligned and the net moment decreases dramatically (Fig. 1), consistent with a global antiferromagnetic (AF) state. This transition is understood to be driven by thermal contraction. Above 154 K, the interplanar (*c*-axis) Mn-Mn distance is greater than 2.85 Å, a critical value below (above) which AF (FM) coupling is commonly observed in this compound family [14]. Surprisingly then, on cooling through 110 K, the system re-enters the FM state which is still non-collinear but with a magnetization easy axis now in the *a-b* plane. The Sm ions, like those of other rare-earths in this family are believed to be trivalent [14]. While ordering among Sm spins has not been seen above 2 K, neutron scattering measurements observe a moment on the Sm sites that grows from 0.3 $\mu_B$ to 0.7 $\mu_B$ on cooling from 110 K to 2 K. [20] The appearance of aligned Sm moments is clearly caused by the Mn exchange field created and not by Sm-Sm interactions.

## Results

In Fig. 1 are shown the real ($\chi'$) and the imaginary ($\chi''$) components of the ac-susceptibility vs. temperature at different frequencies and fields within the reentrant FM phase (data over the full temperature range are in supplemental information [23]). Here, for both applied and ac-fields in the $a$-$b$ plane in which the Sm spins lie, we see below 50 K surprisingly large signals in both $\chi'$ and $\chi''$. While $\chi'$ exhibits a sharp drop at a frequency-dependent temperature, $\chi''$ shows a large frequency-dependent peak at $\omega_0$. This frequency dependence is bounded in temperature between $T \approx 30$ K and $T \approx 45$ K, constrained by the low and high frequency limits of our magnetometer.

The frequency dependence of $\chi_{ac}$ in Fig. 2. shows a step in $\chi'$ accompanied by a peak in $\chi''$ indicating relaxation phenomena with a characteristic energy scale. Besides the temperature dependence of the $\chi''$ peak, shown below to be an activated process, we note the asymmetry between the low and high frequency wings. This suggests that $\chi_{ac}$ is probing two distinct processes, a Debye-like relaxation responsible for a symmetric peak with $\chi''$ vanishing at low frequency and a process in which $\chi''$ is non-zero at low frequency. This latter signal is possibly due to losses caused by irreversible domain wall movement among the Mn spins.

To further understand the relaxation processes, we represent $\chi_{ac}$ in an Argand plot depicting $\chi''$ vs. $\chi'$ in a series of isotherms where frequency varies within each curve (Fig. 2c). For Debye relaxation, i.e. a single relaxation time $\tau = \omega_0^{-1}$ for a given temperature, one expects a semicircle in which $\chi''$ intercepts the $\chi'$ axis at $\chi_T$ and $\chi_S$, the low (isothermal) and high (adiabatic) frequency limits, respectively. We find that the high frequency data coalesce at the left-hand side of the arc with $\chi_S \cong 0.3$ emu/mole but that the curves deviate from each other at lower frequencies, which reflects the non-zero $\chi''$ at low frequency. In addition, the flattening of the $\chi''$- $\chi'$ semicircle indicates a modification of the Debye single-frequency picture to one with a distribution of relaxation times and consequent broadening of the $\chi''$ loss peaks.

### Modeling the frequency dependence

To further parametrize the $\chi'$, $\chi''$ data, we use an empirical generalization of the Debye relaxation model known as the Cole-Cole model [24],

$$\chi_{cc} = \chi'_{cc} - i\chi''_{cc} = \chi_S + \frac{\chi_T - \chi_S}{1+(i\omega\tau)^\alpha} \qquad (1)$$

Where $\chi_{cc}$ represents only the relaxation contribution to $\chi_{ac}$, and $\alpha$ $(0 < \alpha \leq 1)$ represents the width of a symmetrically broadened distribution of relaxation times centered on $\tau$. For $\alpha = 1$ this reduces to the single-$\tau$ Debye model. Fitting to Eq. 1 directly is difficult due to the low frequency constant loss in $\chi''$. To analyze only the Debye process, we fit $\partial\chi'_{cc}/\partial \ln(\omega)$, a method commonly used to isolate relaxation properties in dielectrics [25]. For systems with wide distribution in relaxation times, $(-\pi/2)\partial\chi'_{cc}/\partial \ln(\omega) \approx \chi''$. From our fits (see supplementary information [23]) we find $\alpha$ between 0.52 - 0.60 with an average $\bar{\alpha} \approx 0.55$. The relaxation time distribution is found to be independent of temperature in the measured region and we discuss its origin below.

### Modelling the Temperature Dependence

As mentioned above, the temperature dependence of the $\chi''$ peak frequency is thermally activated and well-characterized by an Arrhenius law,

$$\tau = \tau_0 e^{E_B/k_B T} \qquad (2)$$

where $\tau_0$ is the inverse of the attempt frequency, and $E_B$ the activation energy. We calculated $\tau$ using both the $\chi''$ peak location, $\tau_p = 1/\omega_p$, as well as from the $\partial\chi'_{cc}/\partial\ln(\omega)$ fits, $\tau_{deriv}$. Fits are shown in Fig. 2d, and we find $\tau_{0,p} = 2.2 \pm 0.6$ ps, $E_{B,p} = 869 \pm 10\ k_B$, and $\tau_{0,deriv} = 2.4 \pm 0.7$ ps, $E_{B,deriv} = 838 \pm 11\ k_B$. The difference in $E_B$ between the two methods may be attributed to the FM moment of the Mn sublattice affecting the $\chi''$ signal more than $\chi'$. Both fitting methods give comparable Arrhenius parameters, but we believe that the $\partial\chi'_{cc}/\partial\ln(\omega)$ approach better parameterizes the relaxation process because it addresses only the relaxation from non-Mn spins, which is further borne out in the field dependence discussed below.

The Arrhenius dependence seen in $Dy^{3+}$ SMMs mentioned earlier typically originates from spin-phonon relaxation via two phonons, known as the Orbach process involving low lying states $a$ and $b$ with a state $c$ at a much larger relative energy, $k_B\Delta$. Here a high density of phonons with energy $k_B\Delta$ may induce the transition from $b \to c \to a$ via phonon absorption and emission, expected to be faster than the $b \to a$ direct process [26]. For $\Delta \gg T$, the relaxation time is given by $\tau = (1/\gamma\Delta^3)\exp(\Delta/T)$, where $\gamma$, a constant determined by phonon properties, ranges between $10^3$-$10^5$ $s^{-1}K^{-3}$ for rare-earth ions [27]. This reduces to Eq. 2 for $E_B = \Delta/k_B$ and $\tau_0 = 1/\gamma\Delta^3$. Using $E_B = 838\ k_B$ and $\gamma = 10^3$ $s^{-1}K^{-3}$, we calculate a characteristic relaxation time of 1.7 ps, which is within our measured error estimate for $\tau_0$. Both $Sm^{3+}$ and $Sm^{2+}$ have a first excited state low in energy compared to the other rare-earth ions (approximately 1430 K and 416 K respectively) [27]. These energies are both close to the relaxation energy scale we see in $SmMn_2Ge_2$, suggesting an origin in the Sm free (fire) spin spectrum.

## Field Dependence

The relaxation process persists in dc-fields, $H$, up to saturation values of approximately 1500 Gauss and we thus extended the $\partial\chi'_{cc}/\partial\ln(\omega)$ fitting technique into this field region (see supplemental information for fit parameter dependence [23]). With increasing $H$, corresponding to a reduction in domain wall density in the Mn (ice) sublattice, we observe a shift of the relaxation loss peak to lower temperatures, as well as decreases in $\chi'$, $\chi''$, and $\Delta\chi$ (see Fig. 3), where $\Delta\chi = \chi_T - \chi_S$. In $\chi''$ we observe a decrease in both the magnitude of the loss peak as well as the constant loss at low frequency, resulting in more symmetric peaks. Fig. 3c shows $\Delta\chi$ at 36 K and $\bar{\alpha}$ as a function of dc-field offset. We see that, as $H$ increases, $\bar{\alpha}$ increases to 0.76, approaching the Debye limit. Despite this strong field dependence, $1/\tau_0$ remains in the 0.4 – 2.5 THz range and $E_B$ decreases by only 10% (Fig. 3d), which shows that the *relaxation characteristics are nearly independent of the number of domains*. In the following, we compare our data to other systems that exhibit magnetic relaxation to gain insight into the origin of $E_B$ and examine the implications of our field dependence data.

## Discussion

Relaxation effects such as those exhibited by $SmMn_2Ge_2$ are not common in magnetic systems. They require a spatially-localized, spin-like degree of freedom and an anisotropy field, $H_a$, that sets the attempt frequency, $\tau_0^{-1}$, via the approximation $\tau_0^{-1} \cong g\mu_B H_a/2\pi\hbar$, identifying $\tau_0^{-1}$ with a precession frequency [28]. Such anisotropy is readily achieved in Ising systems, where even in the paramagnetic region, relaxation effects are observed in modest fields, as seen in the antiferromagnet $Dy_3Al_5O_{12}$ [29] and in spin ice compound $Dy_2Ti_2O_7$ [30]. In such cases, the atomic spin is the localized degree of freedom, but in other cases, groups of spins can collectively resonate, examples of which are $\pi$-solitons in the quasi 1D X-Y ferromagnet TMNC ($\omega_0 = 9 \times 10^6$ Hz, $E_B = 5.1\ k_B$) [31], and more generally domain walls in ferromagnets such as $Fe_3O_4$ ($\omega_0 = 1.7 \times 10^8$ Hz, $E_B = 464\ k_B$) [32]. The $E_B$-$\omega_0$ relationships for these systems are compared in Fig. 4.

As already mentioned, Debye-like resonances have also been studied in the SMMs. In Fig. 4 we compare our data to a representative selection of SMMs, including *i*) the dodecanuclear crystalline $Mn_{12}Ac$ [33-35], *ii*) the molecular chain systems [Co(hfac)$_2$NITPhOMe] (CoPhOMe) and [MnF$_4$TPP][TCNE] [36, 37] and *iii*) $Dy^{3+}$ SMMs which show both large attempt frequencies (> GHz) and large effective energy barriers [38] as measured in [Dy(Cp$^{ttt}$)$_2$][B(C$_6$F$_5$)$_4$] [39, 40] and [($\eta^5$-Cp*)Dy($\eta^5$-Cp$^{iPr5}$)][B(C$_6$F$_5$)$_4$] [41]. For completeness, we mention experiments done largely on thin films where relaxation can be characterized as distributions of dwell times in the telegraph noise of charge transport and, like Debye-relaxation measurements, show Arrhenius behavior [42]. These measurements often yield unphysically large (~ $10^{16}$ Hz) values of $\tau_0^{-1}$ (see Fe nano-island data [42]), however, compared to Debye-relaxation measurements, and have been discussed in Ref. [43].

While the Ising barrier is a natural constraint for spin relaxation, *applied* fields can also create such a condition, which is of course the basis for nuclear magnetic resonance. For high-temperature resonance of electronic spins, however, the magnitude of such a field would need to overcome thermal fluctuations, requiring tens of Tesla. An alternative approach is via the half-fire, half-ice concept, where the internal exchange field of the 3*d* elements can create a high-field environment for 4*f* spins residing on a different crystallographic sublattice [12]. We suggest that such a mechanism is operating in $SmMn_2Ge_2$, as discussed next.

As neutron scattering shows, the Sm spins are observable, i.e. polarized, only below the reentrant FM transition and their moment increases in concert with the Mn Bragg widths, implying a substantial interaction between the two sublattices [20]. In other words, the $Sm^{3+}$ moments are polarized by the internal Mn exchange field which, because $Sm^{3+}$ polarization is visible already at 100 K and, because the $Sm^{3+}$ moment is the smallest non-van Vleck moment in the 4*f*'s, implies a substantial exchange field, $> 10^4$ Oe. Also, because $Sm^{3+}$ is known to exhibit isotropic g-factors in various crystal field environments [27], and because Sm-Sm interactions are typically seen at much lower temperatures, one can expect them to behave as paramagnetic spins in a very large Mn-created magnetic field.

Central to the picture proposed below is that the observed behavior is *not* that of relaxing Mn spins. Clearly, individual Mn spins are so strongly coupled to each other that a 1 Oe field would have a negligible effect on their orientation. In addition, while a 1 Oe field is sufficient to displace a domain wall in a soft ferromagnet, domain wall resonances typically display $\tau_0^{-1} \sim 10^6 - 10^8$ s, far lower than our observed $10^{11}$ s values [21, 32]. Finally, we note that such resonances are not observed, to our knowledge, in any soft ferromagnet, rather appearing only when a barrier to domain motion exists, as evidenced by a non-zero coercive field. The coercive fields for $SmMn_2Ge_2$ are $\cong 1$ Oe.

Although the effect is likely not due to Mn spins, we do observe that an applied field greater than 2000 Oe extinguishes the relaxation effect in both $\chi''(\omega = 1/\tau)$ and $\Delta\chi$ (see above). This finding, together with the large $\tau_0^{-1}$ and the size of $E_B$, which are *not* field dependent, strongly suggest that the observed relaxation is due to paramagnetic Sm spins located in Mn domain walls via the following scenario. Domain walls separate regions of anti-aligned spins so within the wall the field varies greatly, and time-dependent changes to the neighboring Mn domains will certainly influence the Sm spins. Such a scenario is supported by the magnitude of the $\chi'$ relaxation signal which corresponds to only $\sim 0.5\%$ of spin-1/2 $Sm^{3+}$ spins per mole, a value consistent with other soft ferromagnets. In Fig. 5 we demonstrate a possible picture for the Sm spin relaxation. For simplicity, we depict the domain walls as sharp and, in zero applied field, as having only a single Sm spin inside. As the ac-field oscillates, Mn spins flip to align with the external field resulting in shifting of domain walls. Paramagnetic Sm spins, originally in the domain wall, align with the newly flipped Mn spins. As the field reaches its maximum or minimum, the Sm spins relax via the much faster Orbach process. This picture is justified by estimating the number of Mn spins flipped during an ac cycle. Assuming an equal number of opposing Mn spins at 0 Gauss, and all Mn spins aligned at $H_{sat}$, then we can approximate the spins flipped from our dc magnetization curves. At a 1 Oe field, we estimate ~ 0.1% of the Mn spins to flip polarization, which is of the same order of magnitude of $Sm^{3+}$ spins involved in the relaxation signal. Furthermore, a distribution of internal fields within a finite-size wall may explain the relaxation time broadening, i.e. $\alpha < 1$, seen in the Cole-Cole analysis. Thus, because domain walls will vanish with applied field on the scale of 1000 Oe, the Sm relaxation signal will also vanish, albeit while retaining its characteristic $\tau_0^{-1}$ and $E_B$,

In the scenario described above, we see how paramagnetic (1/3-fire) spins, when subject to the large internal field due to an ordered (2/3-ice) hetero-spin sublattice can be both *aligned* as well as *switched* by the domain ice sublattice. The above discussion also highlights the reason for neglecting demagnetization effects. Demagnetization corrections are used to calculate the average internal field, which is then assumed to apply to the phenomenon of interest. Here, however, it is likely that our relaxation signal is due to a small subset of spins that experience a field that varies greatly at microscopic distances and is much smaller than the average internal field. In such instances, applying a demagnetization correction would be ill-motivated.

The observation of single-spin-like relaxation in $SmMn_2Ge_2$ may present new options for device design. Unlike many $Dy^{3+}$ SMMs, $SmMn_2Ge_2$ is chemically stable and can be patterned lithographically, allowing domain wall engineering [11]. One caveat is that the temperature where the timescale of magnetic fluctuation coincides with the measurement time (blocking temperature) is found to be 27 K for $SmMn_2Ge_2$, is still too low for applications. Nevertheless, the principle of weakly interacting spins sandwiched between soft FM layers suggests other materials-instantiations of the relaxation behavior presented here especially for thin film growth of heterostructures that mimic $SmMn_2Ge_2$'s unique phase relationships.

Acknowledgments – We thank Chandra Varma, David Lederman, Weiguo Yin, and Dan Dahlberg for useful discussions. This work was supported by U.S. Department of Energy Office of Basic Energy Science, division of Condensed Matter Physics grant DE-SC0017862.

## Figures

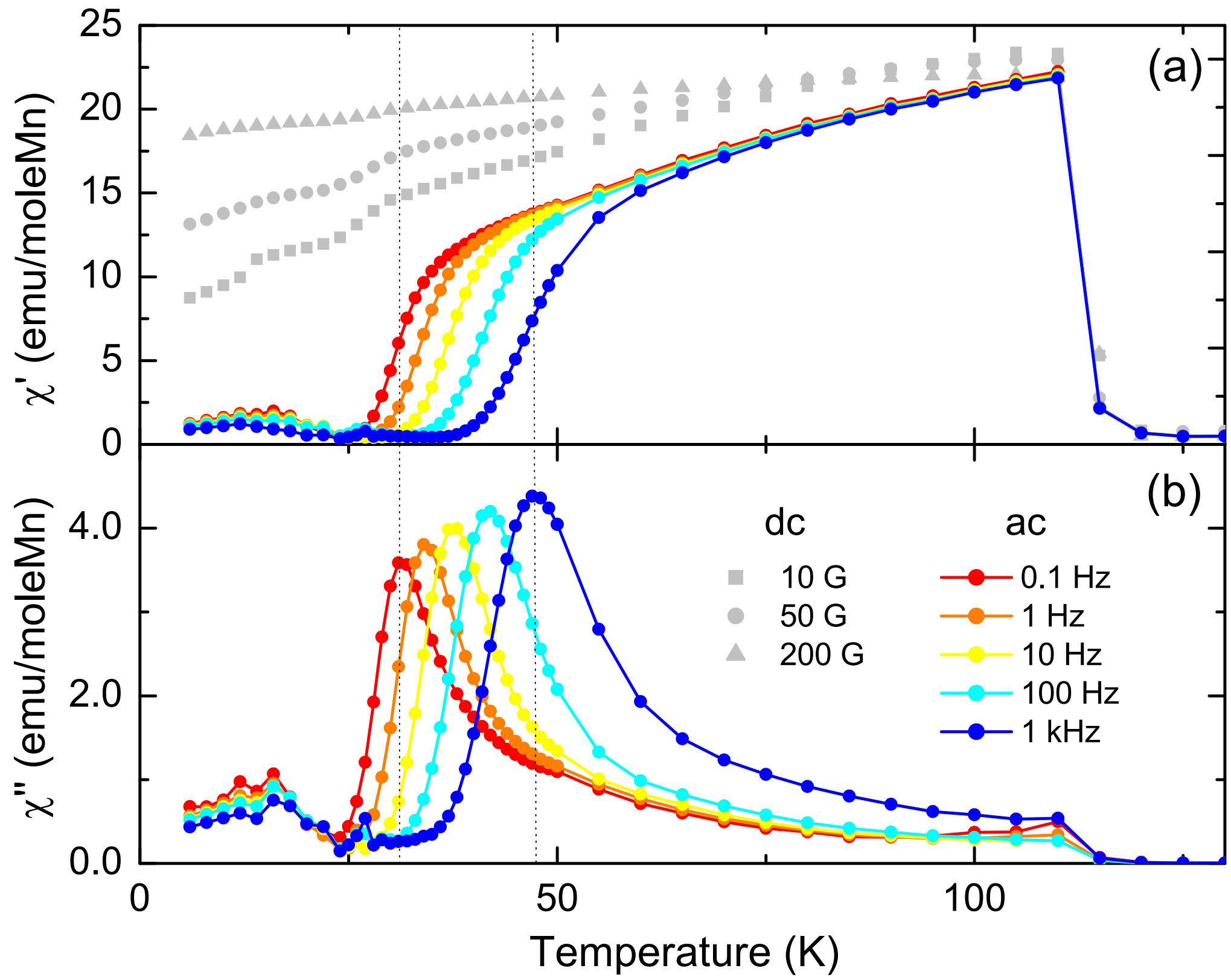

Fig. 1. Ac and dc-susceptibilities below the re-entrant transition with ac and dc-fields perpendicular to *c*-direction. Frequency dependence seen upon cooling below transition with magnetic relaxation from 48 K- 30 K.

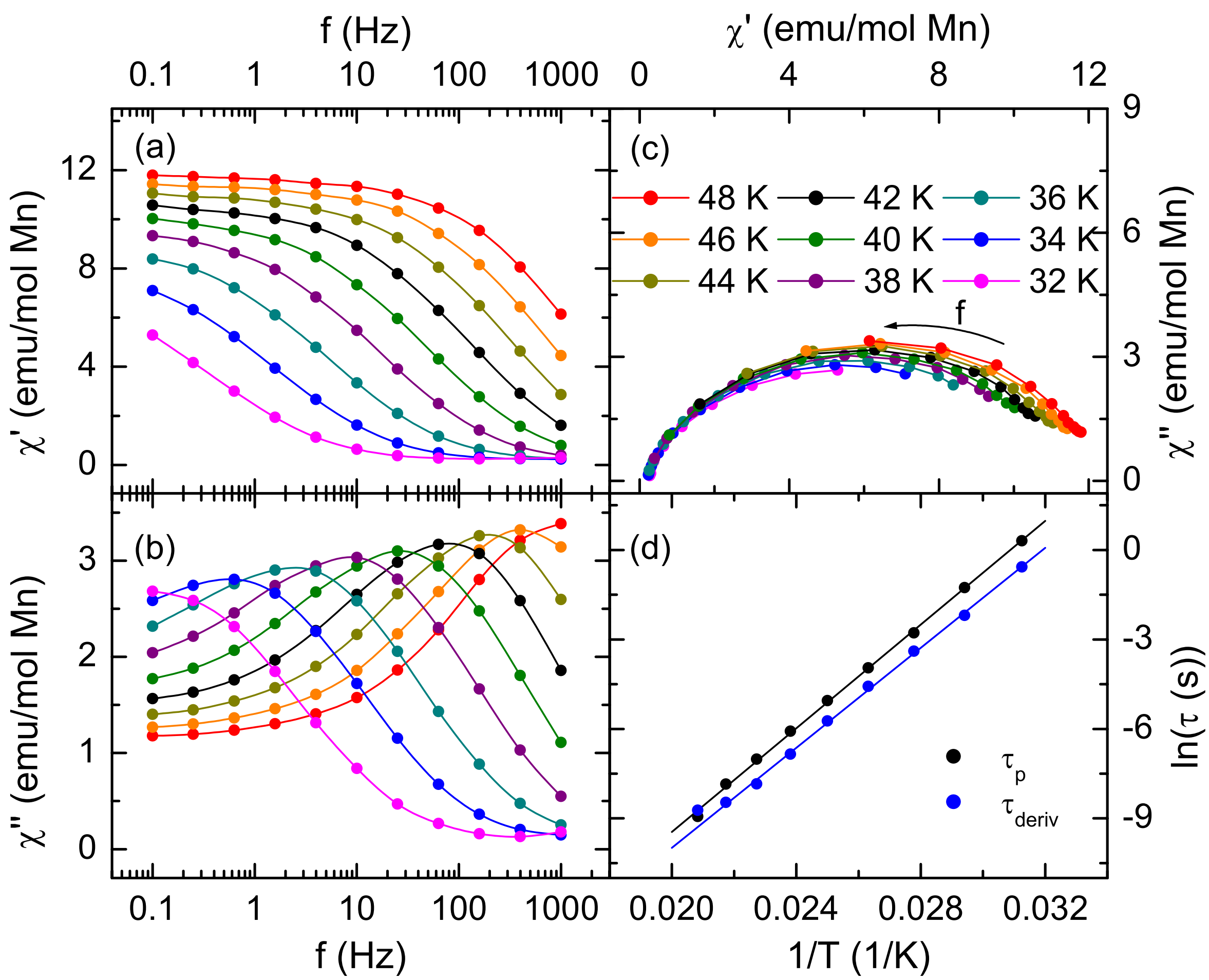

Fig. 2. Ac-susceptibility (real (a) and imaginary (b)) in observable relaxation region. Relaxation shifts to lower frequencies as the sample is cooled. (c) Argand plot of different relaxation isotherms. The distorted semicircles represent a distribution of relaxation times with each point at a different frequency. (d) Arrhenius plot of the mean relaxation time. Two different methods were used to calculate relaxation times, and both show comparable results.

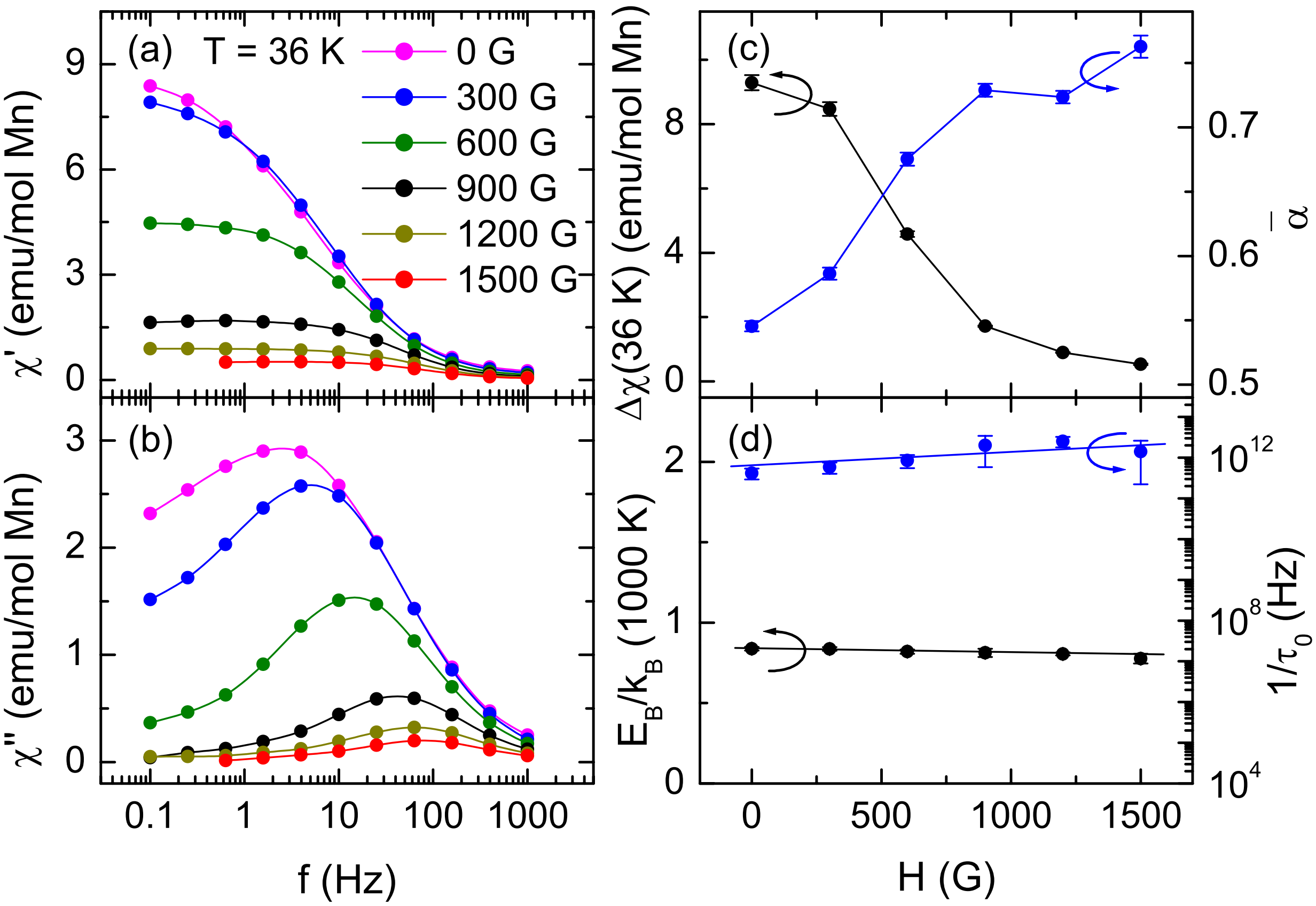

Fig. 3. Dc-field offset dependence of ac-susceptibility (real (a) and imaginary (b)) at 36 K. The lines are to guide the eye. (c) change in Cole-Cole fit parameters $\Delta\chi$ at 36 K (black) and $\bar{\alpha}$ (blue). (d) Dc-field offset dependence of Arrhenius fit parameters $E_B$ (black) and $1/\tau_0$ (blue).

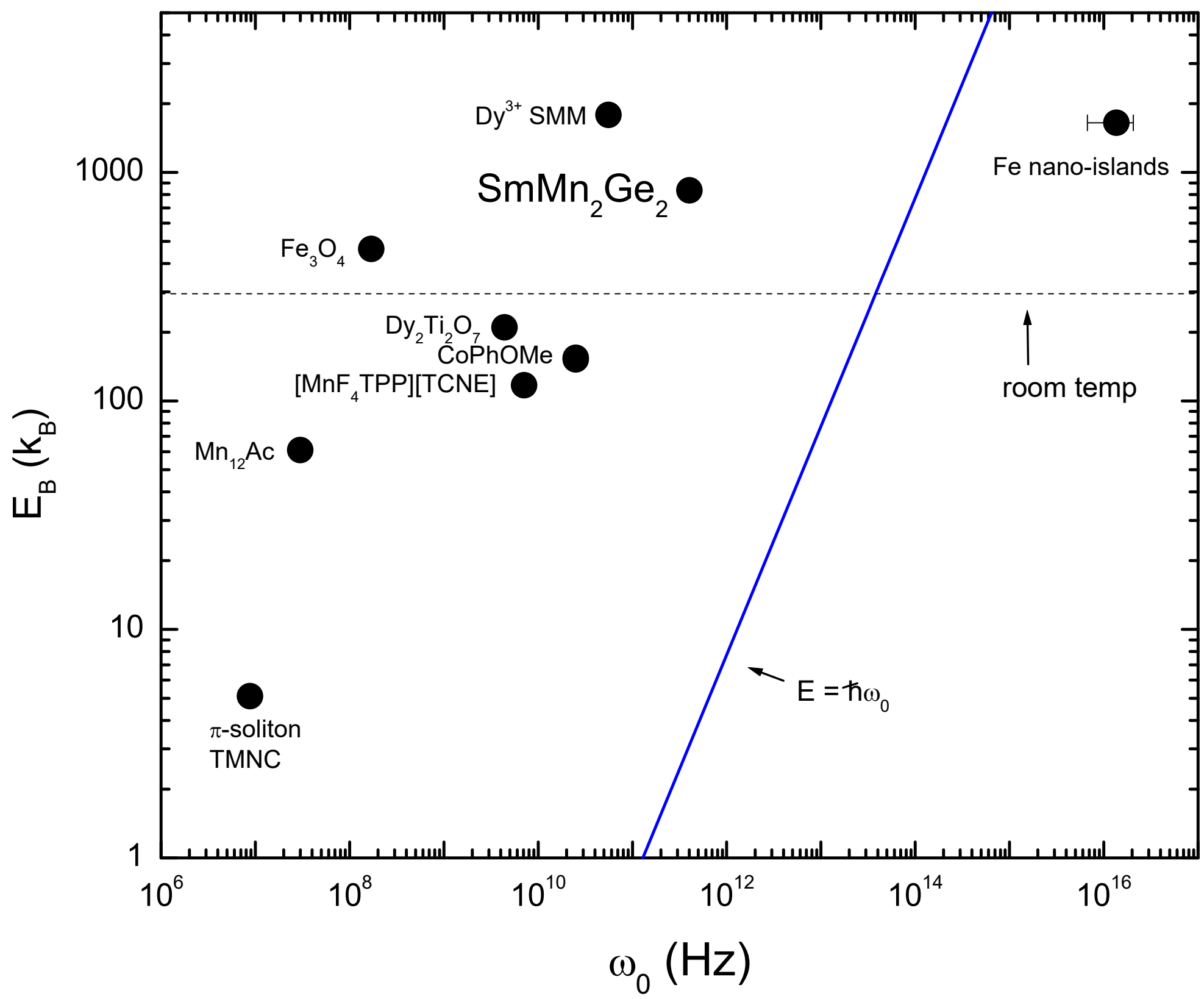

Fig. 4. Comparison of relaxation characteristics between selected single crystal systems, single molecular magnets (SMM), and Fe-nano-island data in the literature. The $Dy^{3+}$ SMM point and the error bars on the Fe nano-island point are representative of a range of relaxation properties. The solid line is the relationship between an isolated two-level system and the frequency of a photon resulting from its relaxation.

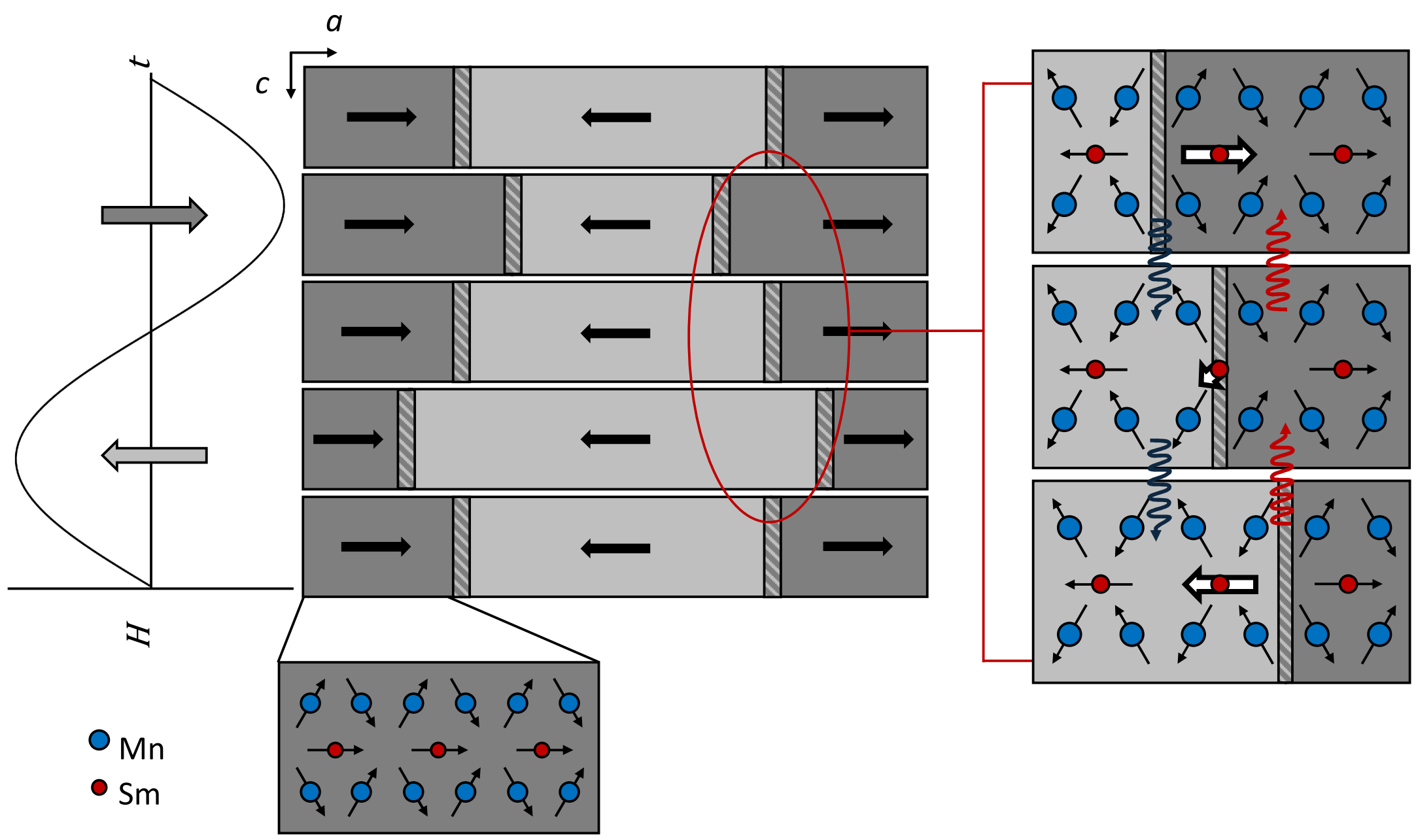

Fig. 5. A possible scenario for relaxation of Sm spins due to movement of Mn domains walls. The domain walls are shown as hatched regions. On the left hand side one sees the growth and shrinking of the Mn domains as the ac-field oscillates. On the right hand side one sees how this domain wall movement flips Sm spins as they align with new Mn polarization. The relaxation is via the Orbach process, depicted as an phonon wave symbol.

Supplementary Note: Quantum Spin Relaxation with THz Attempt Frequency in the 1/3-Fire, 2/3-Ice Ferrimagnet $SmMn_2Ge_2$

M. L. McLanahan and A. P. Ramirez

*Physics Department, University of California Santa Cruz, Santa Cruz, CA, 95064*

**Supplementary Note 1: Dc-Magnetization**

In Fig. S1 $SmMn_2Ge_2$ dc-susceptibility (magnetization divided by field) data are shown for the temperature range 350 K to 2 K. At ~350 K, the Mn ions undergo antiferromagnetic (AF) ordering with a canted ferromagnetic (FM) moment along the $c$-axis. As the system is cooled below 150 K, a pure AF phase is reached. Below 100 K, the FM canted moment re-emerges, now in the $a$-$b$ plane.

**Supplementary Note 2: Ac-Susceptibility**

Ac-susceptibility data taken with a 1.0 Oe drive-field is shown in Fig. S2. Frequency dependence in both $\chi'$ and $\chi''$ is seen below 60 K for drive-fields perpendicular to the $c$-direction. In particular, from 48 K to 32 K, relaxation behavior i.e., a decrease in $\chi'$ with an associated peak in $\chi''$, is seen. However, for drive fields parallel to the $c$-direction in the same temperature range, the change in $\chi'$ is significantly reduced with no associated peaks in $\chi''$. We note that the observed relaxation occurs for drive-fields parallel to the Mn FM canted and Sm moments.

**Supplementary Note 3: Cole-Cole Fitting**

As discussed in the manuscript, we model relaxation in $\chi_{ac}$ using the Cole-Cole model

$$\chi_{CC} = \chi'_{cc} - i\chi''_{cc} = \chi_S + \frac{\Delta\chi}{1 + (i\omega\tau)^{\alpha}}\,, \tag{S1}$$

where $\chi_S = \chi'(\omega \to \infty)$, $\Delta\chi = \chi'(\omega = 0) - \chi_S$, $\tau$ is the inverse of the relaxation frequency, and $\alpha$ lies in the interval $(0,1]$ and accounts for broadening due to a distribution of relaxation times centered on $\tau$. For $\alpha = 1$ Eq. S1 becomes the Debye equation. Like the Debye model, the Cole-Cole generalization requires $\chi''$ vanish for $|\omega - 1/\tau| \gg 0$. In $\chi''$ we observe a low frequency constant offset, which the above model does not account for. Instead, we will use $\chi'_{deriv}$ where

$$\chi'_{deriv} = -\frac{\pi}{2}\frac{\partial \chi'}{\partial \ln(\omega)}. \tag{S2}$$

In Fig. S3a we demonstrate the difference between $\chi'$, $\chi''$, and $\chi'_{deriv}$ at 38 K (middle of temperature region where relaxation is observable). Transforming our data to $\chi'_{deriv}$ results in symmetric peaks as opposed to those seen in $\chi''$. We fit our $\chi'_{deriv}$ data to $-(\pi/2)\partial\chi'_{cc}/\partial\ln(\omega)$ and show results in Fig. S3b.

The temperature and field dependence of the fit parameters is shown in Fig. S4 with uncertainties from the fitting procedure. In the absence of any applied dc-field we find $\alpha$ ranges between $0.517 \pm 0.018$ to $0.596 \pm 0.012$, the variation which may be caused by a combination of a variable fit range and a relaxation signature that moves within this range, a systematic effect not represented by the fit error bars. We thus assume $\alpha$ to be temperature independent, and taking the average, we find $\bar{\alpha} = 0.545 \pm 0.004$. Contrary to this result, we find $\Delta\chi$ decreases as the sample is cooled. The relaxation time, $\tau$, exhibits an exponential temperature dependence, which we present in an Arrhenius plot (see Fig. S4c). We fit the data to Eq. 2 from the manuscript i.e., the Arrhenius equation $\tau = \tau_0 \exp(E_B/k_B T)$, where $\tau_0$ is a characteristic relaxation time, $E_B$ an effective energy barrier, and $k_B$ the Boltzmann constant. In the absence of any applied dc-field we find $\tau_0 = 2.4 \pm 0.7$ ps and $E_B/k_B = 838 \pm 11$ K.

**Supplementary Note 4: Field Dependence**

The above fitting routine was repeated for different applied dc-field offsets. As we increased dc-fields, $\bar{\alpha}$ increased to $0.763 \pm 0.009$, $\Delta\chi$ decreased, and $\tau$ decreased. As before, we fit $\tau$ using an Arrhenius fit and find that $\tau_0$ decreases to $0.4 \pm 0.1$ ps and $E_B/k_B$ decreases to $806 \pm 10$ K at 1200 G. In Fig. S5 we show dc-magnetization curves, as well as $\chi'$ and $\chi''$ for different applied dc-fields. As the field is increased, the magnitude of $\chi'$ and $\chi''$ decrease. Furthermore, we observe a shift of the relaxation peak to lower temperatures.

SUPPLEMENTARY FIGURES

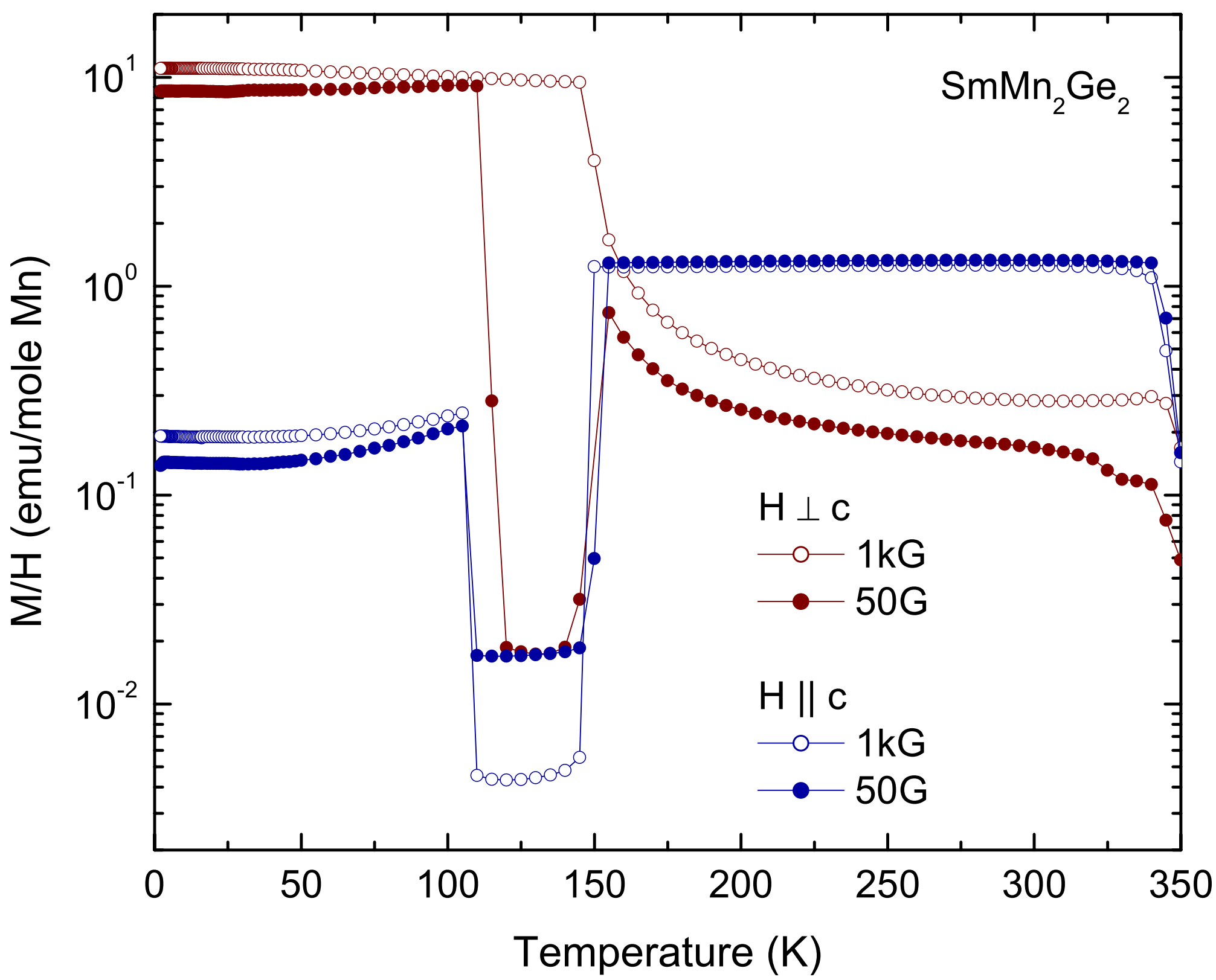

Fig. S1. Dc-magnetization with fields at 50 G and 1 kG aligned perpendicular (red) and parallel (blue) to the *c*-direction.

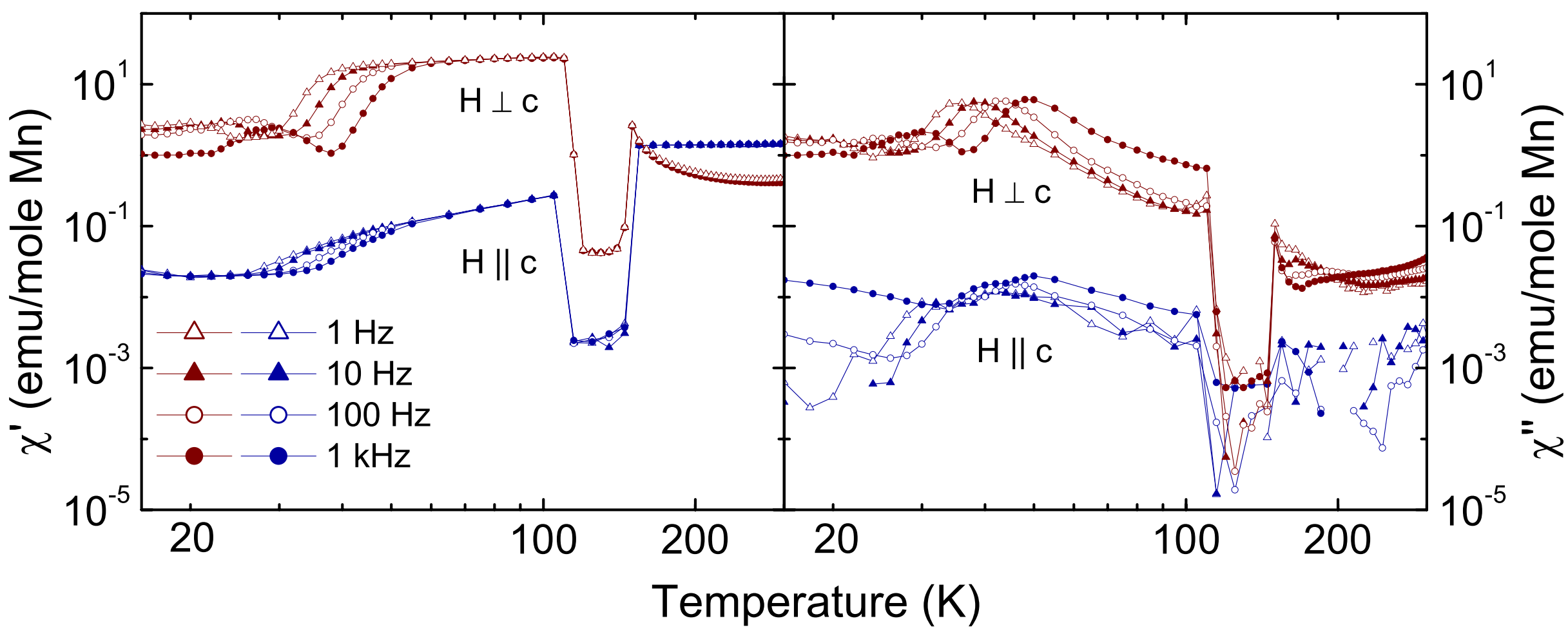

Fig. S2. $\chi_{ac}$ as a function of temperature measured at different frequencies with ac-drive field perpendicular (red) and parallel (blue) to the $c$-direction. Relaxation phenomena seen in $\chi'$ and $\chi''$ when $H \perp c$.

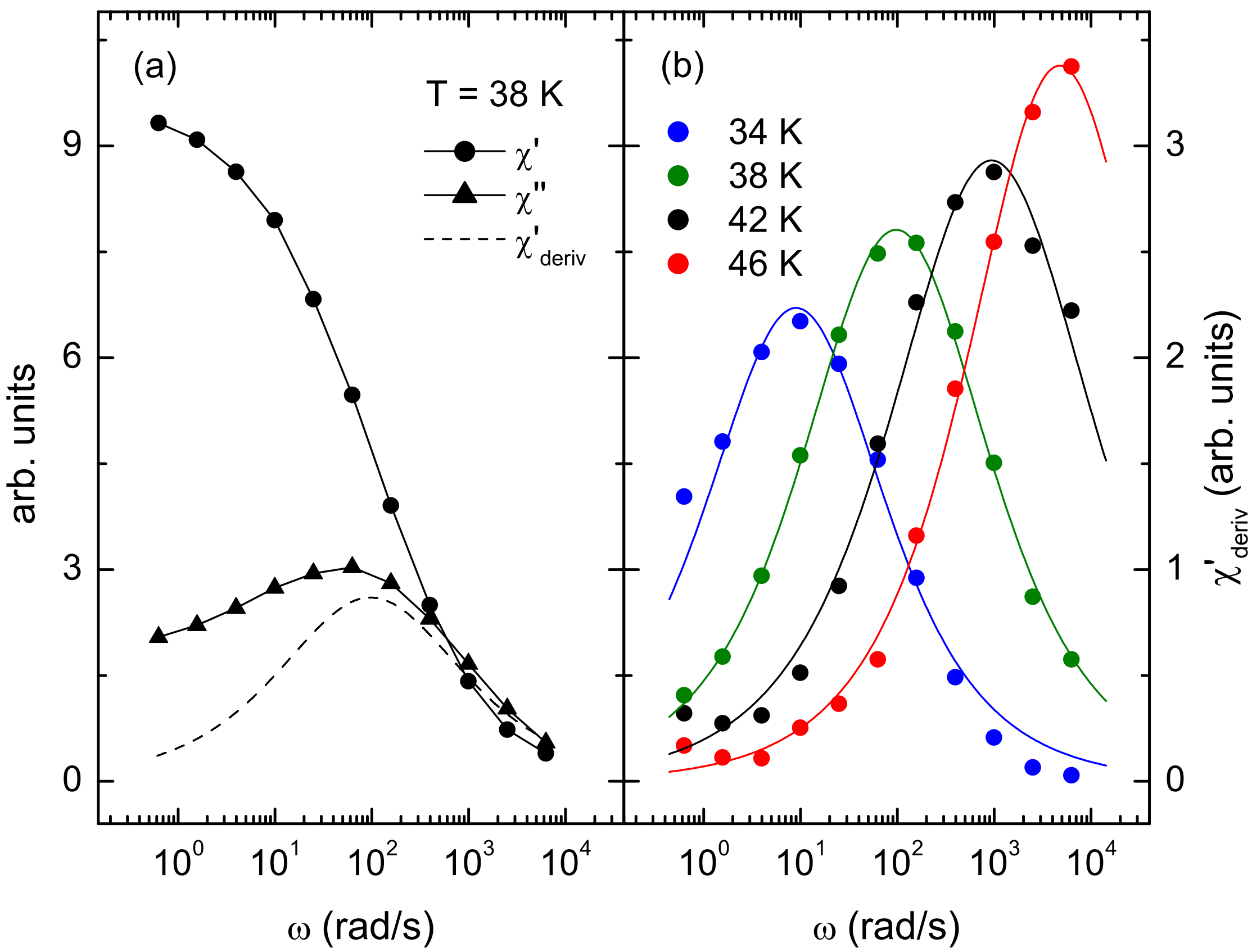

Fig. S3. (a) Comparison of ac-susceptibility and $\chi'_{deriv}$ at 38 K. $\chi''$ is asymmetric due to a low frequency offset, while $\chi'_{deriv}$ has a symmetric peak. (b) $\chi'_{deriv}$ in relaxation region with Cole-Cole fit lines. Ac-drive fields applied perpendicular to $c$-direction.

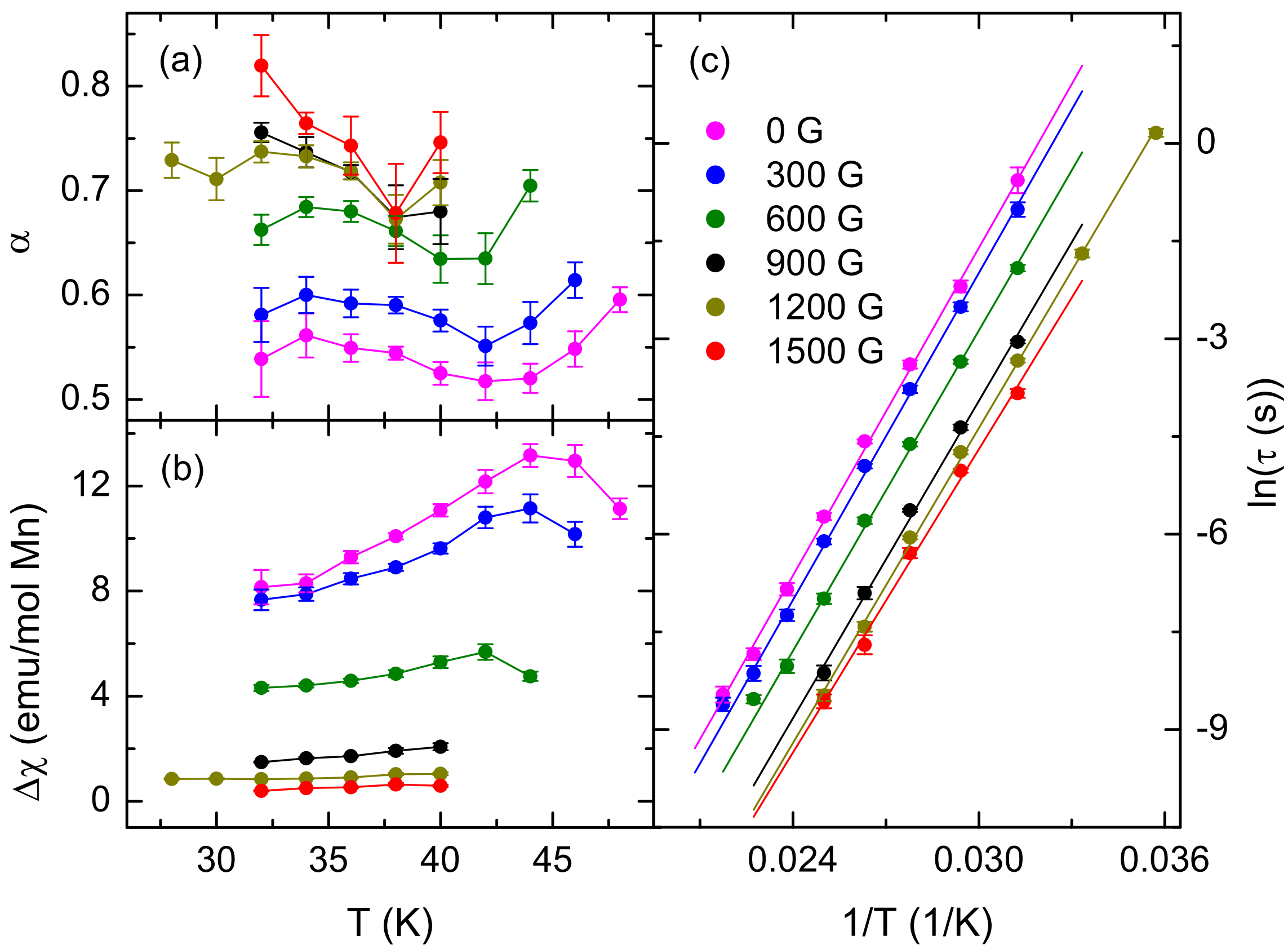

Fig. S4. $\chi_{deriv}$ Cole-Cole fit parameters ($\alpha$ (a) and $\Delta\chi$ (b)) as a function of temperature at different dc-field offsets with lines to guide the eyes. (c) Arrhenius plot of relaxation time $\tau$ with fitting lines. Fields applied perpendicular to $c$-direction.

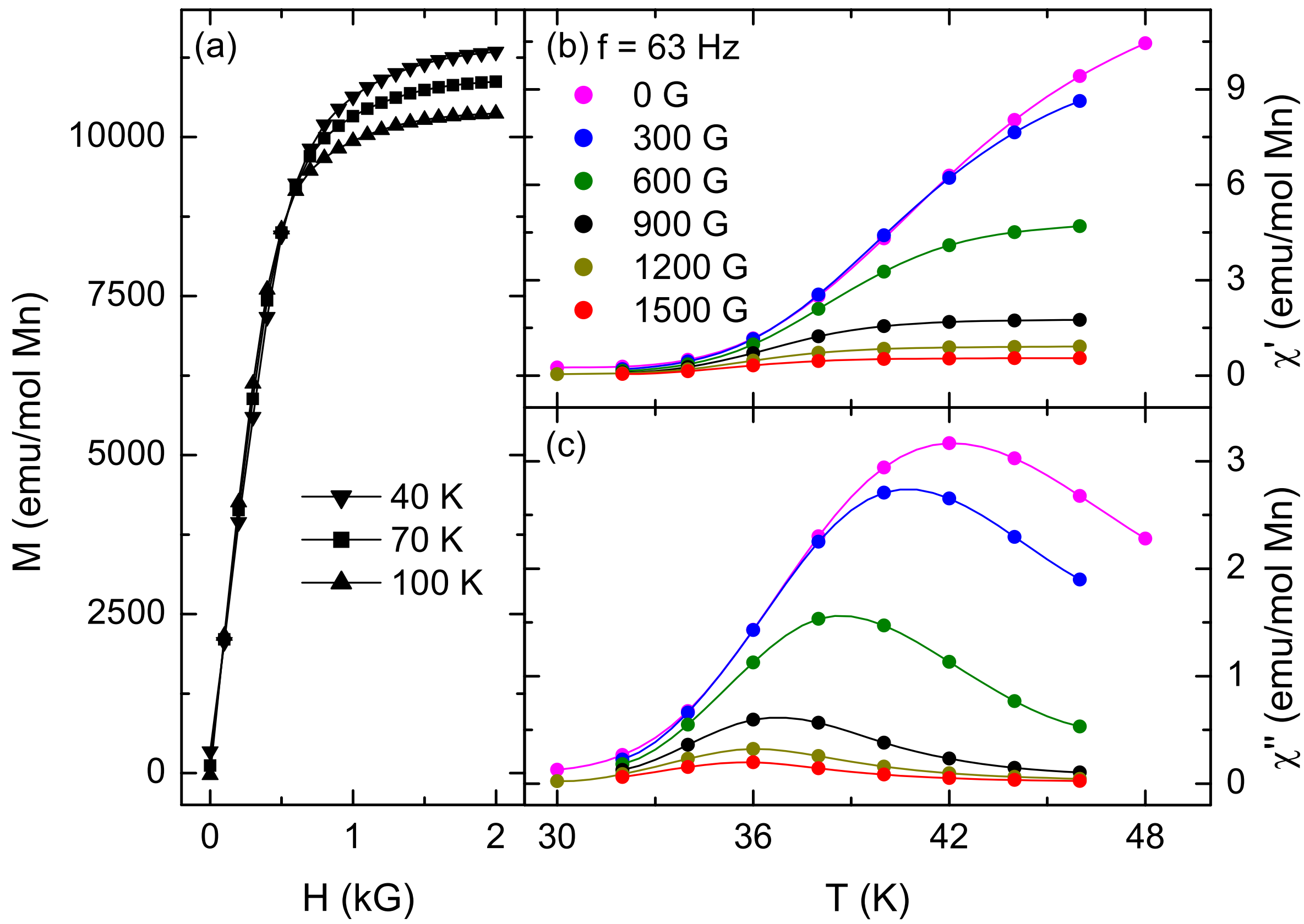

Fig. S5. (a) Magnetization curves at different temperatures in the re-entrant FM phase. Temperature dependence of the ac-susceptibility (real (b) and imaginary (c)) with different dc-field offsets. Data shown for $f = 63$ Hz as a subset of general trend seen. Fields applied perpendicular to $c$-direction.